\documentclass[twocolumn]{aastex631}
\usepackage{amsmath}
\usepackage{comment}
\usepackage{booktabs}
\usepackage[caption=false]{subfig}
\usepackage{savesym}
\savesymbol{tablenum}
\usepackage{siunitx}
\usepackage{soul}
\restoresymbol{SIX}{tablenum}
\usepackage{fontawesome5}

\usepackage{natbib}
\usepackage{verbatim}
\usepackage{mathptmx}  % for better fonts
\usepackage{url}
\usepackage{numprint}
\usepackage{ulem}

\newcommand{\githubrepo}[2]{\href{#1}{#2~\faGithub}}
\newcommand{\IntegratedRun}{\texttt{Integrated Run}}
\newcommand{\IsolatedRun}{\texttt{Isolated Run}}
\newcommand{\Optuna}{O{\scriptsize{PTUNA}}}
\newcommand{\Om}{$\Omega_{\rm m}$}
\newcommand{\se}{$\sigma_{8}$}

\newcommand{\rank}[1]{\text{rk}(#1)}
\newcommand{\NN}[1]{\mathcal{N}}

\definecolor{mycol}{rgb}{0.8, 0.1, 0.1}

\newcommand{\valpm}[2]{#1\,{\scriptstyle \pm\, #2}}
\newcommand{\phantomvalpm}[2]{#1\,\phantom{{\scriptstyle \pm\, #2}}}

\begin{document}

\title{Cosmology with Topological Deep Learning}

\correspondingauthor{Jun-Young Lee}
\email{toti010@snu.ac.kr}

\author[0009-0006-4981-0604]{Jun-Young Lee}
\affiliation{Institute for Data Innovation in Science, Seoul National University, Seoul 08826, Korea}
\affiliation{Center for Theoretical Physics, Department of Physics and Astronomy, Seoul National University, Seoul 08826, Korea}

\author[0000-0002-4816-0455]{Francisco Villaescusa-Navarro} 
\affiliation{Center for Computational Astrophysics, Flatiron Institute, 162 5th Avenue, New York, NY, 10010, USA}
\affiliation{Department of Astrophysical Sciences, Princeton University, 4 Ivy Lane, Princeton, NJ 08544 USA}
 
\begin{abstract}
The standard cosmological model with cold dark matter posits a hierarchical formation of structures. We introduce topological neural networks (TNNs), implemented as message-passing neural networks on higher-order structures, to effectively capture the topological information inherent in these hierarchies that traditional graph neural networks (GNNs) fail to account for. Our approach not only considers the vertices and edges that comprise a graph but also extends to higher-order cells such as tetrahedra, clusters, and hyperedges. This enables message-passing between these heterogeneous structures within a combinatorial complex. Furthermore, our TNNs are designed to conserve the $E(3)$-invariance, which refers to the symmetry arising from invariance against translations, reflections, and rotations. When applied to the Quijote suite, our TNNs achieve a significant reduction in the mean squared error. Compared to our GNNs, which lack higher-order message-passing, ClusterTNNs show improvements of up to 22\% in \Om{} and 34\% in \se{} jointly, while the best FullTNN achieves an improvement of up to 60\% in \se{}. In the context of the CAMELS suite, our models yield results comparable to the current GNN benchmark, albeit with a slight decrease in performance. We emphasize that our topology and symmetry-aware neural networks provide enhanced expressive power in modeling the large-scale structures of our universe.
\end{abstract}

%% Keywords should appear after the \end{abstract} command. 
%% The AAS Journals now uses Unified Astronomy Thesaurus concepts:
%% https://astrothesaurus.org
%% You will be asked to selected these concepts during the submission process
%% but this old "keyword" functionality is maintained in case authors want
%% to include these concepts in their preprints.

%%%%%%%%%%%%%%%%%%%%%%%%%%%%%%%%%%%%%%%%%%%%%%%%%%%%
\section{Introduction} 
\label{sec:intro}
The $\Lambda$CDM model is the best current description of our Universe. This model posits that gravity amplifies nearly scale-invariant tiny quantum fluctuations originating from the primordial cosmos \citep{Harrison1970, Zeldovich1972}. This amplification leads to the formation of cosmological structures we observe today, from galaxy clusters to cosmic voids, collectively referred to as the large-scale structure \citep{Blumenthal1984, DavisLSS, BondLSS}. The model includes a set of parameters that represent fundamental properties of the Universe, such as its expansion rate and the fraction of dark matter and baryons \citep{Riess1998, Spergel2007, Planck2018}.

Determining the value of these parameters is one of the main goals of cosmology. Getting the tightest constraints on these parameters will enhance our understanding of the fundamental physics that governs our Universe. The value of the cosmological parameters influences both the spatial distribution of matter in the Universe and also that of its luminous constituents, such as galaxies. Therefore, studying cosmology by measuring galaxy clustering is a common method used within the discipline \citep{Tinker2012,Alam2017,Ivanov2020,DESY3,DESI2024VI}.

Given a galaxy catalog, which typically contains the three-dimensional positions of galaxies in real or redshift space, cosmologists compress this information into a low-dimensional vector to facilitate analysis. The standard approach involves computing the two-point correlation function, or its Fourier transform, the power spectrum. This method is effective because, on sufficiently large scales or at high redshifts, the power spectrum fully encapsulates the statistical properties of the field, allowing for the complete extraction of underlying information.

Unfortunately, on mildly and non-linear scales, the power spectrum is a suboptimal estimator. In recent years, the community has embarked on a quest to find better summary statistics. Numerous works from different groups have identified multiple statistics that yield tighter constraints on the values of the cosmological parameters than the power spectrum. Among these are higher-order statistics such as bispectrum \citep{Sefusatti2006,Hahn2021MolinoBispectrum,Ivanov2023}, marked power spectrum \citep{Philcox2020,Massara2023}, abundances of voids and clusters \citep{Sahlen2019,Bayer2021}, counts-in-cells \citep{Uhlemann2020}, and wavelets \citep{Allys2020,Valogiannis2022,Eickenberg2022}.

Another approach to solving this problem is to utilize machine learning methods. The concept is to frame the issue as a task of learning the posterior distribution $P(\vec{\theta}|\{x_i\}_{i=i}^N)$, where $\theta$ denotes the vector containing the parameter values and $\{x_i\}_{i=i}^N)$ represents the galaxy catalog itself, instead of employing summary statistics. By doing so, we aim to exhaust the information content embedded in the galaxy catalogs. 

There are various ways of structuring catalog data before feeding it to machine learning algorithms. One possibility is to deposit the galaxies into a 3D grid and then apply computer vision techniques such as convolutional neural networks \citep{Ntampaka2020, Hwang2023, Lesmos2023}. Unfortunately, this approach presents two significant challenges. First, the size of the grid will determine the scales at which the information can be extracted. Second, for a very fine grid (that contains at most one galaxy per voxel), the grid will be dominated by empty voxels because of the sparsity of galaxy catalogs. 

Because of this, the use of geometric deep learning has emerged as a powerful method to tackle this issue. In cosmology, numerous studies have employed point clouds \citep{Anagnostidis2022,DiffusionPointCloud,Chatterjee2024,Lee2024} and graph neural networks \citep[GNNs;][]{CosmoGraphNet, Cranmer2021_allocation_GNN, Makinen2022, Shao2023,deSanti2023,Ho2024_LTU_ILI} to perform parameter inference at the field level. GNNs have also demonstrated effectiveness in various tasks beyond the inference of cosmological parameters, including inferring baryonic or dark matter halo properties of galaxies \citep{Wu2023, PVD2022_HaloGraphNet, PVD2023_WeightMW}, reconstructing the velocity field \citep{Tanimura2024_VelFieldGNN}, predicting intrinsic alignment quantities \citep{Jagvaral2022_IA_GNN}, and enhancing photometric redshifts of galaxies \citep{Beck2019RefinedRR,Tosone2023_photoz_GNN}.  

However, point cloud-based neural networks lack any form of message-passing, and GNNs can only account for pairwise interactions. This limitation makes it challenging for both point cloud-based neural networks and GNNs to effectively address higher-order interactions. For example, it is well known that GNNs struggle with tasks such as counting triangles or measuring the length of the longest cycle \citep{Chen2020_GNN_Triangle, Garg2020_GNN_limitations}.

Higher-order message-passing networks were developed to address this problem \citep{papillon2023architectures}. Numerous studies have explored the power of these neural networks in varying topological domains, including simplicial complexes, cellular complexes, and hypergraphs. Recently, \cite{hajij2023topological} unified these diverse methods within the framework of combinatorial complexes. In this work, we apply these approaches, collectively referred to as topological neural networks (TNNs), to cosmological data for the first time. 

Using two datasets from state-of-the-art N-body and hydrodynamic simulations, we will demonstrate how to generate topological networks on halo and galaxy catalogs by introducing hierarchy between higher-order structures. Next, we will construct our TNNs to preserve the symmetries involved in the problem, thereby ensuring that our models are E(3)-invariant. Thus, we will compare the accuracy achieved by TNNs in performing cosmological parameter inference against the results obtained from GNNs. 

This paper is organized as follows. In Section \ref{sec:data}, we describe the two datasets utilized in this work. In Section \ref{sec:methodology}, we provide a detailed account of topological neural networks, our implementation of $E(3)$-invariance, the training procedure, and the evaluation metrics. The results of our study are presented in Section \ref{sec:results}. Finally, we discuss these results and draw our main conclusions in Section \ref{sec:conclusions}.

\section{\label{sec:data}Data and Benchmarks}
In this section, we outline the details of the data utilized in our study and compare it with the corresponding current benchmarks. For the Quijote suite, we compare our results from topological neural networks with two other benchmarks that employ point clouds and GNNs, respectively \cite{Chatterjee2024} and \cite{MIT_Benchmark}. For the CAMELS suite, we evaluate our results against the GNN model used in \cite{CosmoGraphNet}.

\subsection{QUIJOTE\label{sec:data_quijote}}
The Quijote suite consists of a series of cosmological $N$-body simulations performed using the TreePM {G\scriptsize{ADGET}}-III code, which is an improved version of {G\scriptsize{ADGET}}-II \citep{GadgetII}. These simulations initiate from varied initial conditions at $z=127$ and progress until $z=0$, contained within cubic volumes of $(1~\text{Gpc}/h)^3$. The suite comes in various flavors, not only by varying the standard $\Lambda$CDM cosmological parameters, but also by exploring models that include massive neutrinos, alternative dark energy scenarios, primordial non-Gaussianities, modified gravity, and parity violations \citep{Quijote-PNG, Quijote-ODD}. 

In our analysis, we use the latin-hypercube (LH) set of standard $\Lambda$CDM simulations, executed with a fiducial resolution of $512^{3}$ CDM particles and with fixed cosmological parameters of $\omega=-1$, $M_\nu=0$ eV and $\Omega_k=0$. The LH set consists of 2,000 simulations, with initial conditions produced through second-order Lagrangian perturbation theory. Reflecting its name, cosmological parameters are drawn from specific ranges with varied random seeds to address cosmic variance: $\Omega_m \in [0.1, 0.5]$, $\Omega_b \in [0.03, 0.07]$, $h \in [0.5, 0.9]$, $n_s \in [0.8, 1.2]$, and $\sigma_8 \in [0.6, 1.0]$. From the various available data products, we employ the 5\text{\small,}000 most massive halos identified from the halo catalog using the friends-of-friends (FoF) halo finder.

The dataset used in this study is largely similar to those utilized by the other two benchmarks. The study in \cite{Chatterjee2024} employs the identical LH set, but varies the sampling number of the most massive halos to 1\text{\small,}024, 4\text{\small,}096, and 8\text{\small,}192. We compare our results with their best performing model, which utilizes 8\text{\small,}192 halos. In contrast, \cite{MIT_Benchmark} selects the same 5\text{\small,}000 most massive halos identified by the {\rm R{\scriptsize OCKSTAR}} halo finder \citep{Rockstar}, Another difference is that \cite{MIT_Benchmark} draws from the Big Sobol Sequence set, which includes a total of 32\text{\small,}768 simulations. The difference between the LH and the Big Sobol Sequence lies in the number of simulations, and hence the volume in parameter space. Due to computational constraints, we use the LH set. Given that the other set has a higher density in parameters, we expect to achieve more accurate results if we had trained our models with that dataset. Notably, our model is trained on a smaller number of halos or data samples compared to the other benchmarks.

\subsection{CAMELS\label{sec:data_camels}}
The CAMELS suite comprises both cosmological $N$-body simulations and (magneto-)hydrodynamic simulations \citep{CAMELS_presentation,CAMELS_DR1,CAMELS_DR2}. The fiducial set of (magneto-)hydrodynamic simulations tracks the evolution of $256^{3}$ cold dark matter and gas particles each, starting from $z=127$ and finalizing at $z=0$, in a cubic volume of $(25~\text{Mpc}/h)^3$. In comparison to the Quijote suite, simulations from CAMELS also vary the astrophysical models by including runs from various code groups and subgrid physics models in a smaller box. Since our study focuses on the impact of adding topological analysis rather than the differences between subgrid physics models, we select the IllustrisTNG suite. The simulations from the IllustrisTNG suite are executed by the A{\scriptsize{REPO}} code, which solves magnetohydrodynamics with TreePM on a moving Voronoi mesh \citep{Arepo_Volker2010, Arepo_Weinberger2020}, alongside the same subgrid physics model as for the IllustrisTNG simulations \citep{IllustrisTNG_Weinberger2017,IllustrisTNG_Pillepich2018, IllustrisTNG_Nelson2019}. 

For a direct comparison with the results from \cite{CosmoGraphNet}, we use the same galaxy catalogs from the LH set. Subhalos are identified using the {S\scriptsize{UBFIND}} halo finder, and we apply the selection criteria of a minimum of 20 star particles, corresponding to a stellar mass cut-off of approximately $M_\star \gtrapprox 10^8 \text{M}\odot /h$. The LH set consists of 1,000 simulations, with cosmological parameters fixed at $\Omega_b=0.049$, $\Omega_k=0$, $h=0.6711$, $n_s=0.9624$, $\omega=-1$, and $M_\nu=0$ eV. Other cosmological and astrophysical parameters vary across simulations, with values sampled using the latin-hypercube method: $\Omega_m \in [0.1, 0.5]$, $\sigma_8 \in [0.6, 1.0]$, $A_{\rm SN1}, A_{\rm AGN1} \in [0.25,4.0]$, and $A_{\rm SN2},A_{\rm AGN2} \in [0.5,2.0]$.

\section{\label{sec:methodology}Methodology}

    \begin{figure*}[t]
    \centering
    \includegraphics[width=\textwidth]  {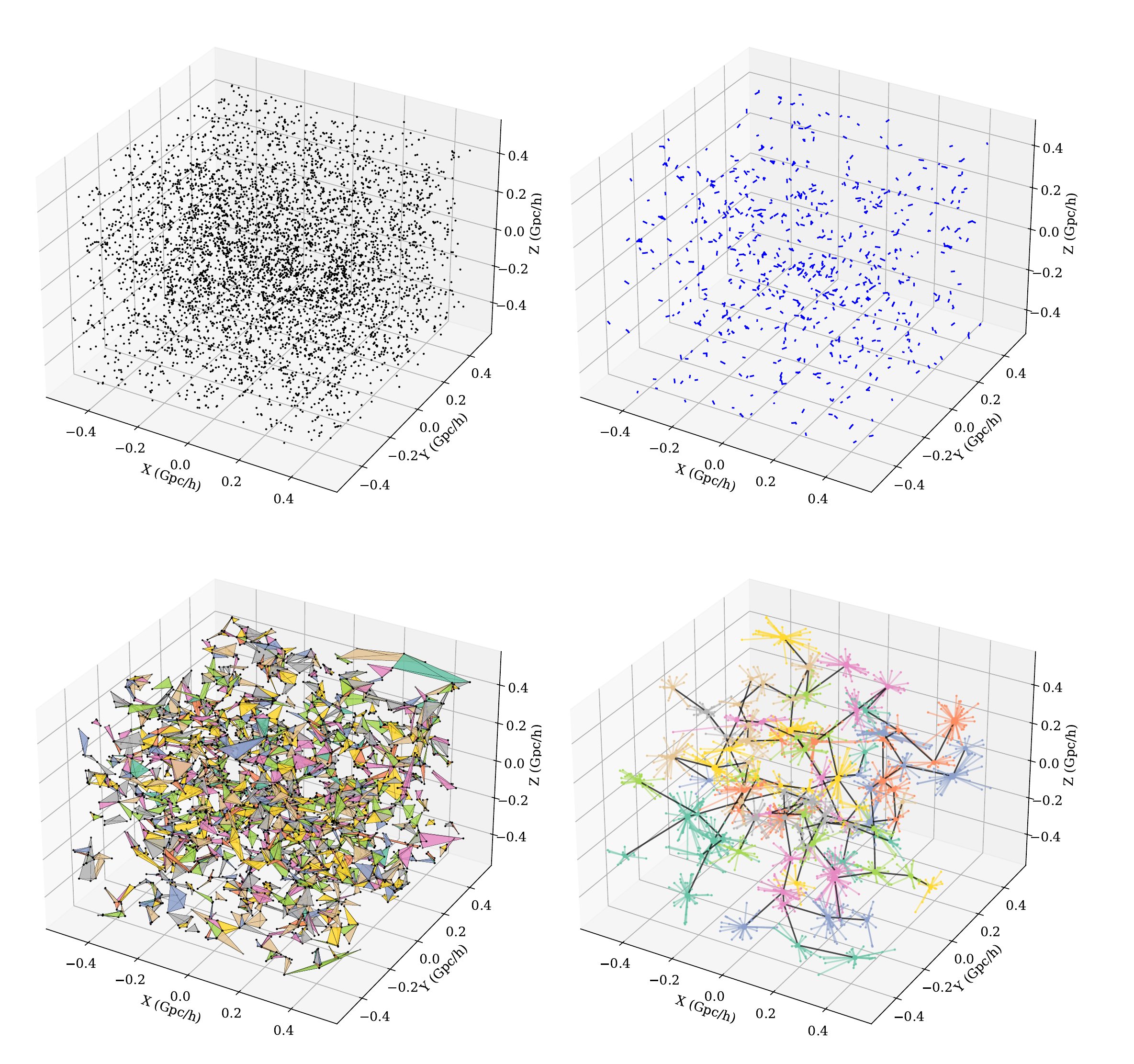}% Here is how to import EPS art
    \caption{\label{fig:structures} Visualization of a combinatorial complex constructed from a halo catalog of a Quijote simulation. Given a halo catalog, the halos represent rank 0 cells, and the collection of halos are referred to as point clouds (top-left panel). Two halos are linked by an edge (rank 1) when their separation is less than $r_{\rm link}$. Creating edges within the point cloud forms a graph (top-right panel). A Delaunay triangulation is performed on the point cloud, and tetrahedra (composed of 4 halos) are identified (bottom-left panel). These tetrahedra represent rank 2 cells. Clusters of tetrahedra (rank 3 cells) are identified using HDBSCAN, and they are shown with colors in the bottom panel. Finally, a minimum-spanning tree is constructed on top of the tetrahedra clusters. Their edges are the rank 4 cells and they are shown as black lines on the bottom-right panel.
    %Visualization of higher-order structures from a sample halo catalog in the Quijote suite --- tetrahedra (\textit{top}) and clusters along with hyperedges (\textit{bottom}). In this illustration, we limit our consideration to the 3\text{\small,}000 smallest-volume tetrahedra. In the \textit{bottom} panel, colored edges illustrate the association of halos with the cluster centers, as determined by Hierarchical Density-Based Spatial Clustering of Applications with Noise (HDBSCAN) applied to tetrahedral cells. The solid black lines indicate connections between clusters, outlining the hyperedges structure formed by linking two separate clusters via the Minimum Spanning Tree (minimum spanning tree) algorithm.
    }
    \vspace{2mm}
    \end{figure*}
    
%\subsection{\label{sec:topological_structures} Extending Sets and Graphs into Topological Structures}
\subsection{Combinatorial complexes}

Point clouds operate on a set of vertices, while graphs introduce a binary or pairwise relationship between two vertices. Topological neural networks work on generalizations of such data structures by incorporating higher-order cells, or structures containing more than two vertices. These networks enable the modeling of higher-order information, which may be difficult to extract with point cloud-based neural networks or GNNs. Furthermore, topological neural networks facilitate effective message-passing over long distances between two remote vertices, which typically require traversal through multiple edges in conventional GNNs, by lifting messages to higher-order cells.

While deep sets and graph neural networks are designed to work on sets and graphs, respectively, combinatorial complexes are the natural input to topological neural networks \citep{hajij2023topological,papillon2023architectures}. A combinatorial complex is defined by the tuple $(S, \chi, \text{rk})$, where $S$ is a set, $\chi \subset P(S) \setminus \{0\}$ is a set of cells, and $\text{rk}:\chi \to \mathbb{Z}_{\geq0}$ is a \textit{order-preserving} rank function.

\begin{figure*}[t]
\centering
\includegraphics[width=0.99\textwidth] {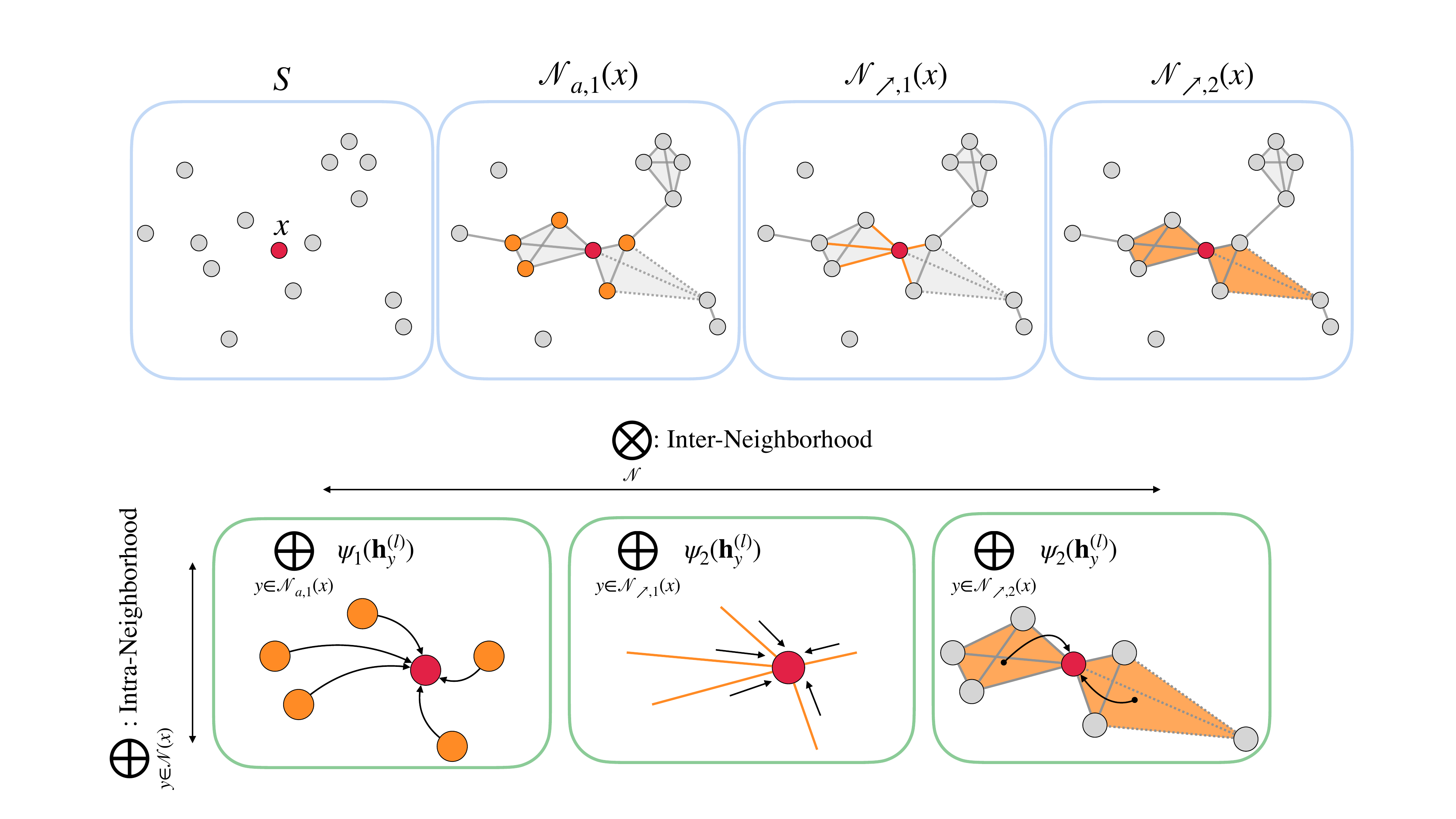}% Here is how to import EPS art
\caption{\label{fig:neighborhoods} This diagram illustrates the identification of neighborhoods within combinatorial complexes and the process of message-passing. The panels highlight a singular red vertex, $x$, alongside its neighboring cells (\textit{orange}), and non-neighboring cells (\textit{gray}). For simplicity, we depict cells of rank 0 (vertices), rank 1 (edges), and rank 3 (tetrahedra). The dotted lines represent disconnected, or combinatorially discarded edges that nonetheless contribute to forming a tetrahedron. The upper panels, from left to right, sequentially display the set $S$, the adjacency neighborhood $\mathcal{N}_{a, 1}(x)$ described by edges, and incidence neighborhoods for edges $\mathcal{N}_{\nearrow, 1}(x)$, as well as for tetrahedra $\mathcal{N}_{a, 2}(x)$. The lower panels illustrate the message-passing scheme as established in Equation \ref{eqn:message-passing}. Each panel demonstrates the aggregation process within individual neighborhoods (intra-neighborhood aggregation), and ultimately, the aggregation process across all defined neighborhoods (inter-neighborhood aggregation) is shown.}
\end{figure*}

To formally define such higher-order networks, we begin with a non-empty set $S$. Subsets of $S$ with a cardinality exceeding two are termed higher-order cells, while a two-element set is referred to as an edge cell, and a singleton set is known as a vertex cell. We now define a series of concepts needed to work with topological neural networks:

\begin{itemize}
\item \textbf{Cells}. Combinatorial complexes are made up of cells of different ranks. Formally, given a set $S$ a cell is defined as $x \in P(S) \setminus \{0\}$. For instance, vertices are singletons and are typically assigned rank 0. Edges contain two vertices and typically are assigned with rank 1. Cells containing two or more vertices are called higher-order cells and their rank is typically set to $\geq2$. 

\item \textbf{Rank}. The different cells in a combinatorial complex are assigned a rank that creates a hierarchy among the cells. The rank function, $\text{rk}$, assigns the same or higher rank to cell $y$ than to cell $x$ if $x \subseteq y$ for all $x, y \in \chi$. Specifically, $\quad \forall s \in S, \, \{ s \} \in \chi$ and if $x, y \in \chi \text{ and } x \subseteq y, \text{ then } \rank{x} \leq \rank{y}$. 
This rank function allows us to establish hierarchies among cells by grouping those with an identical rank $k$ into $\chi^{k} \subset \chi$, or \textit{$k$-cells}. Individual vertices will naturally be assigned the rank $k=0$. We note that the rank is not determined by the number of elements or cardinality. For example, vertices can be grouped into rank 2 cells using some clustering criterion, and each of these cells will have a different number of vertices.

\item \textbf{Neighborhood function}. A neighborhood function defines the neighbor cells of a given cell: $\mathcal{N}:S \to P(P(S))$ where $P(S)$ denotes the power set of $S$. Neighborhood functions are defined so that the pair $(S, \mathcal{N})$ forms a topological space. This definition of neighbors greatly introduces the flexibility in the data structures that the neural network can process. For example, we can define a hypergraph in the set $S$, with hyperedges being elements of $\chi \subset P(S) \setminus \{0\}$. Unlike normal graphs, hypergraphs can have edges that are defined by three or more vertices.

\end{itemize}

%Naturally, faces on graphs can be defined if the edges and vertices form a cycle comprising three or more of each. Here, edges that border a face are typically presumed or inherently understood. Alternatively, the higher-order network allows for a more relaxed approach by combinatorially omitting certain edges while still maintaining hierarchical relationships among them. These relations with additional flexibility, in that the presence of one relation does not imply another, are referred to as \textit{set-type relations}. Hypergraphs exemplify the extreme of this property. 

\subsection{\label{sec:message-passing}Message-passing over combinatoral complexes}

In GNNs, messages are exchanged between vertices and edges. In topological neural networks, the data is structured as a collection of cells of different ranks, and those cells can exchange messages with neighboring cells. We note that, given a cell, neighbor cells do not need to have the same dimension as the cell itself.

In this section, we explain how the different neighborhoods are defined for cells in TNNs and how messages are passed between cells. Given a combinatorial complex $(S, \chi, \text{rk})$, we can define the following neighborhoods:

\begin{enumerate}
    \item \textbf{Incidence Neighborhood}:  
    Cells are said to be incident if for $x, y \in \chi, \quad x \subsetneq y \text{ or } y \subsetneq x.$
    We can define the \(k\)-up incidence neighborhood function as
    \begin{eqnarray}
    \mathcal{N}_{\nearrow,k}(x) = \{ y \in \chi \mid x \subsetneq y, \, \rank{y} = \rank{x} + k \}
    \end{eqnarray}
    and the $k$-down incidence neighborhood function as
    \begin{eqnarray}
    \mathcal{N}_{\searrow,k}(x) = \{ y \in \chi \mid x \subsetneq y, \, \rank{y} = \rank{x} - k \}
    \end{eqnarray}
    
    \item \textbf{Adjacency Neighborhood}:  
    The adjacency neighborhood of a cell $x \in \chi$ consists of all cells $y \in \chi$ with the same rank, that are incident to a common higher-order cell. Formally we define a $k$-adjacency neighborhood function
    \begin{eqnarray}
    \mathcal{N}_{\text{a},k}(x) = \{y \in \chi \mid \rank{x} = \rank{y}, \exists z \in \chi \text{ s.t. } \nonumber \\
    \rank{z} = \rank{x} + k, \text{ and } x,y \subsetneq z\}
    \end{eqnarray}
    
    \item \textbf{Coadjacency Neighborhood}:  
    The coadjacency neighborhood of a cell $x \in \chi$ consists of all cells $y \in \chi$ with the same rank, that hosts a common lower-order cell. Formally we define a $k$-coadjacency neighborhood function
    \begin{eqnarray}
    \mathcal{N}_{\text{co},k}(x) = \{y \in \chi \mid \rank{x} = \rank{y}, \exists z \in \chi \text{ s.t. } \nonumber \\
    \rank{z} = \rank{x} - k, \text{ with } z \subsetneq x \text{ and } z \subsetneq y\}.
    \end{eqnarray}
\end{enumerate}

Utilizing incidence, one can calculate the neighborhood matrices for two separate set of cells designated with different ranks. Conversely, using (co)-adjacencies allows for the organization of neighborhood matrices of cells among the cells with the same rank.

Using these neighborhood matrices as pipes to pass messages, we must define the data itself tied to each cell. Mathematically, the $k$-\textit{cochain space}, denoted as $\mathcal{C}^k(\chi, \mathbb{R}^d)=\{ H_k: \chi^k \to \mathbb{R}^d\}$ is the data vector associated with cells of rank $k$ and is denoted as $k$-\textit{cochain}. More directly, the data vectors of $k$-cells can be identified in this simplistic form: $H_k = [h_{x^{k}_{1}}, ...,h_{x^{k}_{n}}]$ with $n=|\chi^{k}|$. For traditional GNNs, 0-cochains are the data associated with the vertices, while 1-cochains represent the edge features. We emphasize that cells of different ranks will, in general, have cochains of different dimensions.

Now, we have all the building blocks for designing a topological neural network on a combinatorial complex. Our last step is to formulate the actual message-passing using the neighborhood matrices and the cochains. We use the convolutional operators for \textit{push-forward} and \textit{merge node} operations as follows:

\begin{enumerate}
    \item \textbf{Convolutional Push-Forward}:  
    The push-forward is a map that transforms an \(i\)-cochain into a \(j\)-cochain, denoted as 
    \begin{eqnarray}
    \mathcal{F}_{G;W} : \mathcal{C}^{i}(\chi, \mathbb{R}^{s_{\rm in}}) \to \mathcal{C}^{j}(\chi, \mathbb{R}^{t_{\rm out}})
    \end{eqnarray}
    In practice, given an $i$-cochain $H_i$ of size \(|\chi_i| \times s_{\rm in}\), we define a trainable weight matrix \(W\) of size \(s_{\rm in} \times t_{\rm out}\),  a cochain map $G:\mathcal{C}^{i}(\chi) \to \mathcal{C}^{j}(\chi)$ of size \(|\chi_j| \times |\chi_i|\), which together enable the transformation to a $j$-cochain of size \(|\chi_j| \times t_{\rm out}\) through the following operation,
    \begin{eqnarray} \label{eqn:pf}
    H_i \to K_j = G * H_{i}W.
    \end{eqnarray}
    Here, using Einstein's summation convention, we define the operation $A * B = \bigoplus_{l}A_{kl}B_{lm},$
    where $\bigoplus_{l}$ denotes a permutation-invariant, intra-neighborhood aggregation function for all cells $z_l \in \chi^{j}$ neighboring $x_k \in \chi^{i}$. Note that neighborhood matrices can naturally become good candidates for $G$. 
    
    \item \textbf{Convolutional Merge Node}:  
    Given the definition of convolutional push-forwards, we define a merge node operation that completes the higher-order message-passing operation. Naturally extending the definition above we define
    $
    \mathcal{M}_{\mathbf{G;W}} : \mathcal{C}^{i_1} \times \mathcal{C}^{i_2} \times \cdots \times \mathcal{C}^{i_n} \to \mathcal{C}^{j}
    $ as
    \begin{eqnarray} \label{eqn:merge_node}
    \mathcal{M}_{\mathbf{G;W}} &= \beta \left( 
    \bigotimes_{k=1}^n \mathcal{F}_{G_k;W_k}(H_{i_k})
    \right) \nonumber \\
    &= \beta \left( 
    \bigotimes_{k=1}^n G_k * H_{i_k} W_k
    \right)
    \end{eqnarray}
   where $\bigotimes_{k=1}^n$ serves as an inter-neighborhood aggregation function applied to all neighborhood functions $\mathcal{N}_1, \cdots, \mathcal{N}_n$, $\beta$ denotes an activation function, $\mathbf{G}=(G_1,\cdots, G_n)$ represents a tuple of cochain maps, and $\mathbf{W}=(W_1,\cdots, W_n)$ signifies a tuple of learnable weights. 

\end{enumerate}

The idea of higher-order message-passing is based on the above two operations. Using a neighborhood matrix for $G$, message-passing is naturally enabled between the $i$-th and $j$-th cells of different ranks, connected by the non-zero elements of $G_{ij}$. 

We can summarize the above steps into the following equation, which describes how a cell $x$, with features $\mathbf{h}_x^{(l)}$ at layer $l$, is updated:
\begin{eqnarray} \label{eqn:message-passing}
    \mathbf{h}_{x}^{(l+1)} = \mathbf{h}_{x}^{(l)} + \beta\left[\bigotimes_{k=1}^{n}\bigoplus_{y\in \mathcal{N}_{k}(x)} \psi_{\mathcal{N}_{k}, \rank{x}}(\mathbf{h}_{y}^{(l)})\right]~.
\end{eqnarray}
Let us break it down to explain each component. First, we consider all neighbor cells of rank $k$ to cell $x$: $y\in \mathcal{N}_{k}(x)$. Each of those cells will have features denoted as $\mathbf{h}_{y}^{(l)}$. We take these cell features and pass them through a non-linear, learnable function denoted by $\psi_{\mathcal{N}_{k}, \rank{x}}(\cdot)$. Next, we perform a permutation-invariant intra-neighborhood aggregation operation $\bigoplus_{y\in \mathcal{N}_{k}(x)}$ to all elements in the neighborhood. The procedure above is then repeated for all cells of different ranks and the results are aggregated: represented as $\bigotimes_{k=1}^{n}$. Finally, the result is passed through a non-linear activation function $\beta$, and the cell features are updated by summing the value obtained from all the above operations with the cell features. We note that the equation above models how cell features are updated in a vanilla TNN. The actual equation we will use for our TNNs is slightly different to enforce E(3)-invariance and to make our model more expressive. We describe this in detail in Section \ref{sec:symmetry}.

Figure \ref{fig:neighborhoods} illustrates the neighborhoods and the message-passing scheme defined in Equation \ref{eqn:message-passing} on an example combinatorial complex. The top-left panel depicts a set $S$, or a point cloud, which lacks higher-order structures. We focus on a particular cell of interest: vertex $x$, highlighted in red. In contrast, the other three top panels exhibit a combinatorial complex that includes cells such as edges and tetrahedra with ranks 1 and 2. We identify three distinct neighborhoods for $x$. Cells that are neighbors are depicted in orange, whereas non-neighbors are colored gray. The adjacency neighborhood $\mathcal{N}_{a,1}(x)$ comprises vertices that are connected via edges. The incidence neighborhoods $\mathcal{N}_{\nearrow,1}(x)$ and $\mathcal{N}_{\nearrow,2}(x)$ each include edges and tetrahedra associated with $x$. The three bottom panels describe how messages are sent from neighboring cells. The intra-neighborhood aggregation, denoted in $\bigoplus_{y \in \mathcal{N}(x)}$, aggregates the messages originating from neighboring cells identified from a single neighborhood function. Finally, we perform the inter-neighborhood aggregation operation denoted in $\bigotimes_{\mathcal{N}}$, over the defined collection of neighborhood functions. 

In practice, we build these higher-order message-passing neural networks using T{\scriptsize{OPO}}X \citep{hajij2024topox}, incorporating aggregation operations on sparse neighborhood matrices ($\bigoplus_l$ in Equation \ref{eqn:pf}) with \texttt{pytorch-sparse}.

\subsection{\label{sec:topology_cosmic_data}Building Topologies on Cosmological Data}
Given the fact that the cosmic web is a complex 3D structure characterized by halos, filaments, voids, sheets...etc., one may wonder whether a topological representation of the data, instead of a graph, may ease the task of extracting cosmological information from it. With TNNs, we can potentially model not only individual galaxies but also clusters of galaxies or even filamentary structures of the cosmic web. Furthermore, TNNs inherently incorporate messages from higher-order cells involving more than two halos or galaxies, allowing for the capture of higher-order correlations, unlike the two-point interactions of GNNs. Additionally, efficient long-range message-passing can enhance the representational capacity of our architectures, especially on larger scales.

To construct and train a topological neural network, it is essential to establish combinatorial complexes on the cosmological data. Forming the hierarchy in combinatorial complexes demands a highly increased level of ad hoc engineering compared to point clouds or graphs. %This process resembles associating a binary relation (an edge) where their separation is less than a specified linking length, $r_{\rm link}$, used as a hyperparameter for optimization \cite[see][]{CosmoGraphNet}. It is noteworthy to point out that there are many variations in expressing cosmological data, but we opt for the following architecture to be explained.
Given a set point cloud $\{x_i\}_{i=1}^N$ where $x_i\in\mathbb{R}^d$, we construct a combinatorial complex as follows:
\begin{itemize}
\item \textbf{Rank 0}. The individual halos/galaxies represent rank 0 cells. We also denote these elements as vertices.
\item \textbf{Rank 1}. We create an edge between two halos/galaxies if they are within a distance $r_{\rm link}$. 
These cells represent connections between two rank 0 cells. They correspond to the traditional edges in GNNs. We also call these elements edges.
\item \textbf{Rank 2}. Given the point cloud, we perform a Delaunay triangulation. From it, we can identify tetrahedra composed of 4 halos/galaxies. These tetrahedra represent our rank 2 cells. We refer to these elements as tetrahedra.
\item \textbf{Rank 3}. The identified tetrahedra do not sample the volume uniformly, but they cluster in a similar manner as halos/galaxies in the cosmic web. We identify tetrahedra clusters by applying the Hierarchical Density-Based Spatial Clustering of Applications with Noise \citep[HDBSCAN;][]{HDBSCAN} algorithm to the tetrahedral centroids. We refer to these tetrahedra clusters as rank 3 cells. We denote these cells as clusters.
\item \textbf{Rank 4}. Finally, we can connect the rank 3 cells through a minimum spanning tree (minimum spanning tree). We denote the edges of this graph as rank 4 cells. We refer to these cells as hyperedges.
\end{itemize}

%In our hierarchical framework, we establish five distinct ranks of cosmic structure. At the zeroth rank lie the individual halos or galaxies. The first rank comprises edges within the linking radius ($r_{\rm link}$), while the second rank consists of tetrahedra constructed through Delaunay triangulation. We then derive the third rank by applying the Hierarchical Density-Based Spatial Clustering of Applications with Noise (HDBSCAN) algorithm to the tetrahedral centroids, forming distinct clusters. The fourth and highest rank introduces hyperedges, constructed by connecting the clusters via a minimum spanning tree (minimum spanning tree) algorithm. These hyperedges cells serve as potential pathways for cluster-level information exchange. 

Figure \ref{fig:structures} shows a combinatorial complex constructed using an example simulation of the Quijote suite. The top-left panel delineates individual halos, whereas the top-right panel illustrates the edges between pairs of halos. The bottom-left panel displays the rank 2 cells (tetrahedra) in different colors, which potentially capture additional small-scale information through three- and four-point correlations not accessible to the two-point interactions modeled in GNNs. Finally, the bottom-right panel illustrates the structure at larger scales, showing tetrahedra clusters connected by black solid lines (minimum spanning tree edges) that form hyperedge cells. The striking visual difference between these two panels manifests how the topological structure of the cosmic web can be better described by the hierarchy of cells rather than by the graph itself.

    \begin{figure*}[t]
    \centering
    \includegraphics[width=0.99\textwidth]  {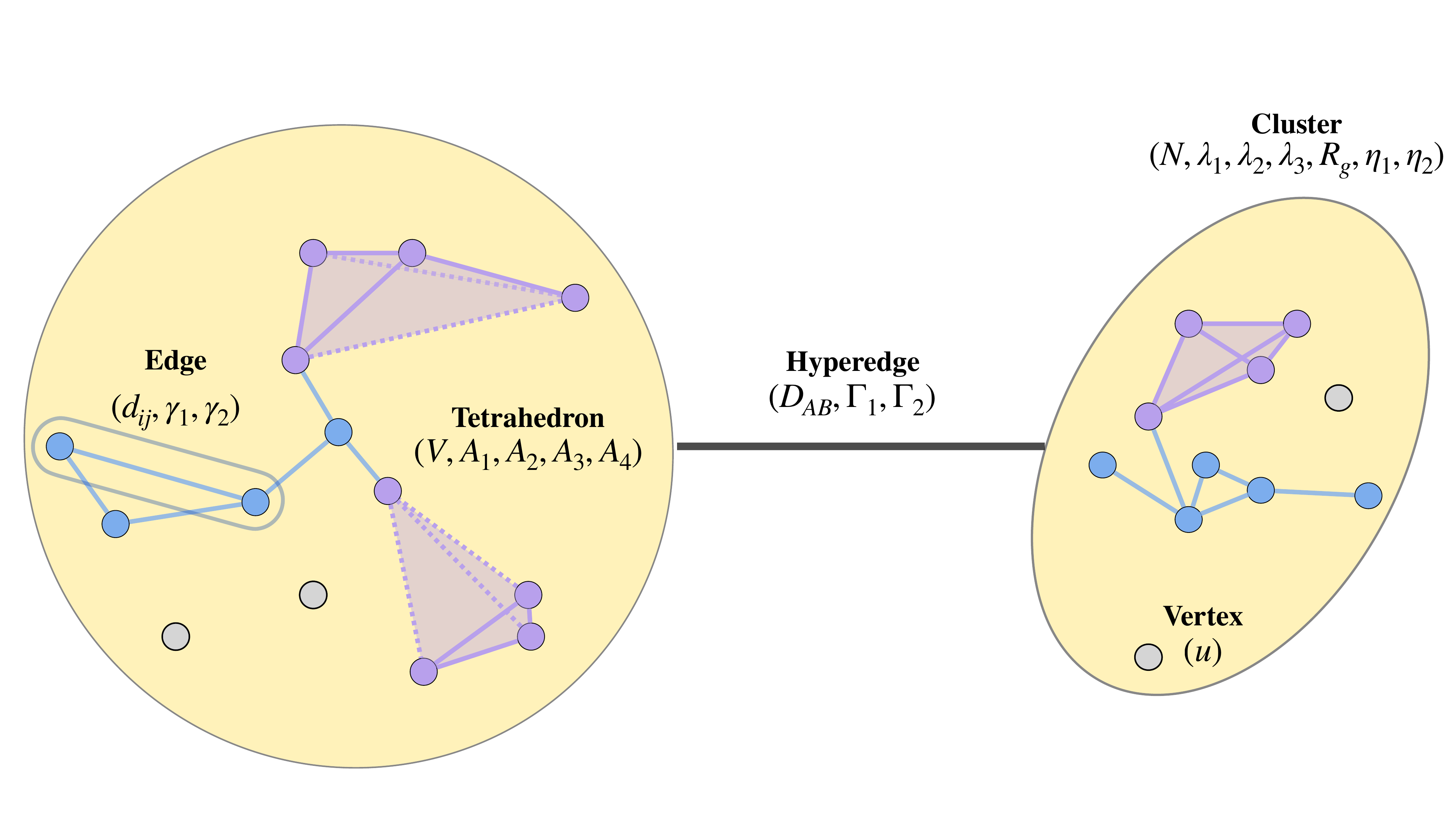}% Here is how to import EPS art
    \caption{\label{fig:cells} This diagram shows different cells of a combinatorial complex and their associated E(3) features. The combinatorial complex exhibits vertices (rank 0) in gray, edges (rank 1) in \textit{blue}, tetrahedra (rank 2) in \textit{purple}, clusters of tetrahedra (rank 3) in \textit{yellow}, and hyperedges (rank 4) in \textit{black solid line}. Vertices and edges are colored based on the highest rank of the cells out of vertices, edges, and tetrahedra they are incident to, with clusters and hyperedges excluded for visual clarity. For instance, \textit{blue} vertices indicate incidence to edges but not to tetrahedra. Note that certain edges forming the tetrahedron may be excluded if the distance $d_{ij}>r_{\rm link}$, indicating the combinatorial complex's structural flexibility as shown in \textit{dotted purple lines}. Refer to Section \ref{sec:symmetry} for more information on the definition of assigned scalars.}
    \vspace{2mm}
    \end{figure*}

    \begin{figure*}[t]
    \centering
    \includegraphics[width=0.99\textwidth]  {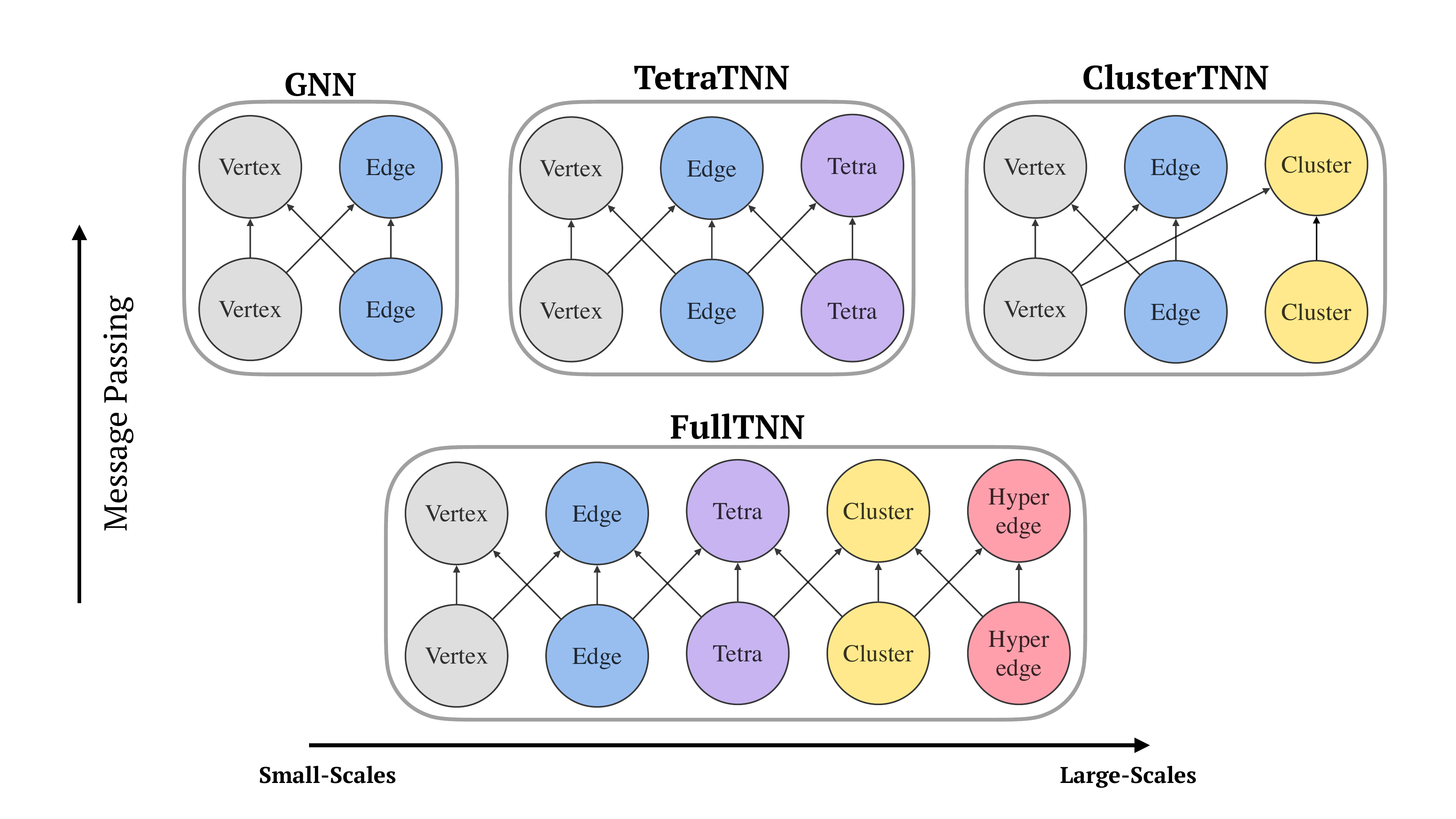}% Here is how to import EPS art
    \caption{\label{fig:architecture} Depicted are tensor diagrams illustrating the structures of the neural networks utilized in this research: the GNN layer (\textit{top left}), the TetraTNN layer (\textit{top center}), the ClusterTNN layer (\textit{top right}), and the FullTNN layer (\textit{bottom}). These layers may be stacked vertically. The circles denote the $k$-cells representing the collection of all cells of rank $k$ within the combinatorial complex. Each of the $k$-cells are color-coded: vertices (\textit{gray}), edges (\textit{blue}), tetrahedra (\textit{purple}), clusters (\textit{yellow}), and hyperedges (\textit{red}). The directed arrows linking different $k$-cells symbolize the convolutional push-forward operation, detailed in Equation \ref{eqn:pf_cci}. The confluence of various arrows represents the convolutional merge node operation as defined in Equation \ref{eqn:merge_node}.}
    \vspace{2mm}
    \end{figure*}

\subsection{\label{sec:symmetry}Symmetries and Geometric Invariants}

The task we want to perform in this study is parameter inference from halo/galaxy catalogs in real-space. Thus, it would be desirable that our model would be: 1) permutational invariant, 2) translational invariant, 3) rotational invariant, and 4) invariant under parity transformations (reflections). We thus construct our model to be both permutational and E(3) invariant. We follow recent studies that have shown how incorporating equivariance or invariance serves as powerful geometric information, especially in higher dimensions \citep{EGNN,SEGNN, EMPSN}. 

In our case, we construct combinatorial complexes whose cells contain $E(3)$-invariant scalar features such as lengths and angles, which naturally leads the messages passed between neighbors to be $E(3)$-invariant. Figure \ref{fig:cells} displays the geometric quantities assigned to different ranks of cells. The features we use for the different cells are as follows. 
\begin{itemize}
    \item Rank 0: $u$. Vertices have a scalar quantity associated to them that we set by sampling a uniform distribution from 0 to 1. We do this mostly because our networks are very powerful and tend to overfit the data very easily. The rank 0 features are designed as a model regularizer. We have explicitly checked that if we do not use features for rank 0 cells, our results do not change, although training is more difficult.
    
    \item Rank 1: $(d_{ij}, \alpha_{ij}, \beta_{ij})$. For the edges, we follow the approaches of \cite{CosmoGraphNet}\footnote{Details on preprocessing simulations into graphs are provided in \cite{SW_CosmographNet} and available on their \githubrepo{https://github.com/PabloVD/CosmoGraphNet}{Github repository}.} and \cite{UniversalScalars}. Specifically, for an edge connecting vertices with positions $\mathbf{x}_{i}$ and $\mathbf{x}_{j}$, we compute their Euclidean distance as $\mathbf{d}_{ij}=\mathbf{x}_{i}-\mathbf{x}_{j}$. Then the edge features are constructed as $d_{ij}=\|\mathbf{d}_{ij}\|$,
$\gamma_{1}=\mathbf{x}_{i}\cdot\mathbf{x}_{j}/(\|\mathbf{x}_i\| \|\mathbf{x}_j\|)$, and $\gamma_{2}=\mathbf{d}_{ij}\cdot\mathbf{x}_{i}/(\|\mathbf{x}_i\|d_{ij})$.

    \item Rank 2: $(V, A_1, A_2, A_3, A_4)$. For tetrahedra, we use the volume $V$ and the area of the four faces $A_i$ as cell features.
    
    \item Rank 3: $(N, \lambda_1, \lambda_2, \lambda_3, R_g, \eta_1, \eta_2)$. A rank 3 cell represents a cluster of $N$ tetrahedra. From the centroid positions of the tetrahedra, $\mathbf{c_i}$, we compute the covariance matrix $\text{Cov} = \frac{1}{N} \sum_{i=1}^N (\mathbf{c}_i - \mathbf{c})(\mathbf{c}_i - \mathbf{c})^T$ where $\mathbf{c}=\frac{1}{N}\sum_{i=1}^N \mathbf{c}_i$ is the centroid of the cluster. We then calculate the eigenvalues $\lambda_1, \lambda_2, \lambda_3$ in descending order of magnitude and the gyradius $R_g=\sqrt{\frac{1}{N} \sum_{i=1}^N \left\|\mathbf{c}_i - \mathbf{c}\right\|^2}$. Finally, we select the two most significant eigenvectors, corresponding to the major and intermediate axes of cluster covariance, $\mathbf{e}_1, \mathbf{e}_2$, and define angles $\eta_i=\mathbf{e_i} \cdot\mathbf{c}/(\|\mathbf{e}\| \|\mathbf{c}\|)$. Thus, the scalar properties associated with each rank 3 cell are $(N, \lambda_1, \lambda_2, \lambda_3, R_g, \eta_1, \eta_2)$.
    
    \item Rank 4: $(D_{AB}, \Gamma_{1}, \Gamma_{2})$. A rank 4 cell represents the connection (edges between two rank 3 cells (clusters) A and B. Thus, we can use the same scalar properties as used for the rank 1 cells (edges), that is, the distance ($D_{AB}$) and angles ($\Gamma_1, \Gamma_2$) between them. 
    
\end{itemize}
We note that all distances, areas, and volumes are scaled by $r_{\rm link}$, $r_{\rm link}^2$, and $r_{\rm link}^3$, respectively.

Given the above cell features, one can construct messages between cells $X$ and $Y$ of the form $m_{x,y}=\psi\left(\text{Inv}(\{\mathbf{x}_{x}\}_{x \in X}), \text{Inv}(\{\mathbf{x}_{y}\}_{y \in Y}) \right)$ where $\psi$ is an arbitrary update function and $\text{Inv}(\cdot)$ represent the cell scalar (invariant) quantities defined above. However, this type of message is just a subset of all possible invariant classes. In order to make our model more expressive, we follow \cite{ETNN} and use messages with invariant quantities defined over the neighborhood of heterogeneous cells: $\psi\left(\text{Inv}(\{\mathbf{x}_{x}\}_{x \in X}, \{\mathbf{x}_{y}\}_{y \in Y})\right)$. We note that there are multiple ways of deriving scalars involving two heterogeneous cells, but we employ the Euclidean distance between centroids,
\begin{eqnarray} \label{eqn:euclidean}
    d_{E}\left(X,Y\right) = \left\| \frac{1}{N_X}\sum_{x\in X}\mathbf{x}_{x} -\frac{1}{N_Y}\sum_{y\in Y}\mathbf{x}_{y} \right\|
\end{eqnarray}
and the Hausdorff distances as follows,
\begin{eqnarray} \label{eqn:hausdorff}
    d_H(X, Y)=\max \left[ \max_{x \in X} \min_{y \in Y} \| \mathbf{x}_{x}{-}\mathbf{x}_{y} \|, \max_{y \in Y} \min_{x \in X} \| \mathbf{x}_{y}{-}\mathbf{x}_{x} \| \right]
\end{eqnarray}

We develop a matrix of scalars $D \in M_{n_1,n_2}(\mathbb{R})$, defined such that $D_{ij} = d(X_i, Y_j) \odot N^{k_1,k_2}_{ij}$. In this context, $n_1 = |\chi^{k_1}|$, $n_2 = |\chi^{k_2}|$, with $X_i \in \chi^{k_1}$, $Y_j \in \chi^{k_2}$, and $N^{k_1, k_2}$ representing the neighborhood matrix between the $k_1$ and $k_2$-cells. Consequently, an element of the matrix $D$ is non-zero only if the respective cells are neighboring, as we perform an element-wise multiplication between the distance and the neighborhood matrix element. The distance function $d$ may be selected from either Euclidean or Hausdorff metrics, and ranks $k_1$ and $k_2$ can be identical or distinct. Utilizing the convolutional push-forward operation as defined in Equation \ref{eqn:pf}, we can add messages generated from these cell-to-cell invariants as follows, 
\begin{eqnarray} \label{eqn:pf_cci}
    K_j = (G * H_{i}W_1) + (D * H_{i}W_2)
\end{eqnarray}
where $W_1, W_2$ are the learnable weights. Equation \ref{eqn:pf_cci} states the augmented convolutional operation used to generate any kind of $E(3)$-invariant message.

Again, at the individual cell level, the message $\mathbf{m}_{x}^{(l)}$ to be passed to cell $x$ in layer $l$ is defined very similarly to Equation \ref{eqn:message-passing}.
\begin{eqnarray} \label{eqn:message-passing_cci}
    \mathbf{m}_{x}^{(l)} = \beta\left[\bigotimes_{k=1}^{n}\bigoplus_{y\in \mathcal{N}_{k}(x)} \psi_{\mathcal{N}_{k}, \rank{x}}(\mathbf{h}_{y}^{(l)}, \text{Inv}(\mathbf{x}_x, \mathbf{x}_y))\right]
\end{eqnarray}
We incorporate the $E(3)$-invariant, defined as $\text{Inv}(\mathbf{x}_x, \mathbf{x}_y) = d(\mathbf{x}_x, \mathbf{x}_y)$. The feature vector of cell $x$ at layer $l$ is updated via $\mathbf{h}_{x}^{(l+1)} = \mathbf{m}_{x}^{(l)} + \mathbf{h}_{x}^{(l)}$.

\subsection{\label{sec:architecture}Neural Network Architecture}
With all the building blocks of message-passing between heterogeneous cells, we can construct layers and, ultimately, complete architectures. Figure \ref{fig:architecture} illustrates the \textit{ tensor diagrams} of the message-passing layers designed in this study. Individual arrows represent the convolution operation in Equation \ref{eqn:pf_cci}, while their confluence signifies the merge node operation defined in Equation \ref{eqn:merge_node}. 

Given the different cells of combinatorial complexes, we can construct various layers that will update the features of these cells. In this paper, we examine four distinct layers (see Fig. \ref{fig:architecture}):
\begin{itemize}
\item \textbf{GNN}. In this case, only cells of rank 0 and 1 are updated. These layers represent the standard layers in GNNs.

\item \textbf{TetraTNN}. This layer considers cells of rank 0, 1, and 2 and updates their features by passing messages among cells whose ranks only differ by one at most.

\item \textbf{ClusterTNN}. This layer considers cells of rank 0, 1, and 3 and updates their feature properties according to the message routes defined in the top-right panel of Fig. \ref{fig:architecture}. 

\item \textbf{FullTNN}. This layer uses all cells in a combinatorial complex and updates their features by exchanging messages with neighbor cells whose rank differ by one at most.
\end{itemize}

A TNN can be constructed by stacking these layers. For instance, a traditional GNN can be formed by stacking the GNN layers defined previously. In general, a TNN does not always require the same layers. For example, one can combine FullTNN with GNN and TetraTNN to construct a TNN. While this approach is feasible, we do not adopt this strategy in this work. Instead, we create TNNs by stacking layers of the same type. The number of layers we stack is a hyperparameter that we optimize.

After the last layer, we perform a global pooling:
\begin{equation}
\xi=\bigoplus_{\mathbf{x}_i\in \chi^{k}}\mathbf{x}_i
\end{equation}
where $\chi^{k}$ represents the $k$-cells, encompassing all the cells of a specific rank $k$. For GNNs\footnote{Note that our GNNs are not completely identical to do GNNs is \cite{CosmoGraphNet} or \cite{MIT_Benchmark}. Our GNN is rather close to the Equivariant GNN (EGNN) used in \cite{MIT_Benchmark} except that we only use invariance and add invariant scalars arising from heterogeneous node-edge pairs.} and TetraTNN, the pooling is performed over the rank 0 cells, while for ClusterTNN and FullTNN we pool over rank 3 and all rank cells, respectively. We then concatenate that vector with $[N_0, N_1, N_2, N_3]$, where $N_i$ is the number of cells with rank $i$ and $N_4$ is neglected due to the nature of minimum spanning tree, $N_4=N_3-1$. The resulting vector is finally passed through a multi-layer perceptron to return the mean and standard deviation of the considered parameters. We refer the reader to our \githubrepo{https://github.com/Byeol-Haneul/CosmoTopo}{Github repository} to explore the details of our neural network.

In this study, we test the performance of different architectures to observe the effectiveness of adding topological information to traditional GNNs. We divide the training runs into two flavors: 
\begin{itemize}
\item \textbf{Integrated Run}. In these models, all four layers participate as a tunable hyperparameter. However, this might cause the optimizer to be biased towards rather simplistic architectures that easily reach higher performances with small trials of tuning other hyperparameters. We note that in this case, while the layer type is a hyperparameter, the model is constructed with a single type. In other words, a TNN will always be built using the same type of layer, not mixing layers.
\item \textbf{Isolated Run}. In this case, the TNN is composed only of layers of a given type. 
\end{itemize}

    \begin{figure*}[t]
    \centering
    \includegraphics[width=0.99\textwidth]  {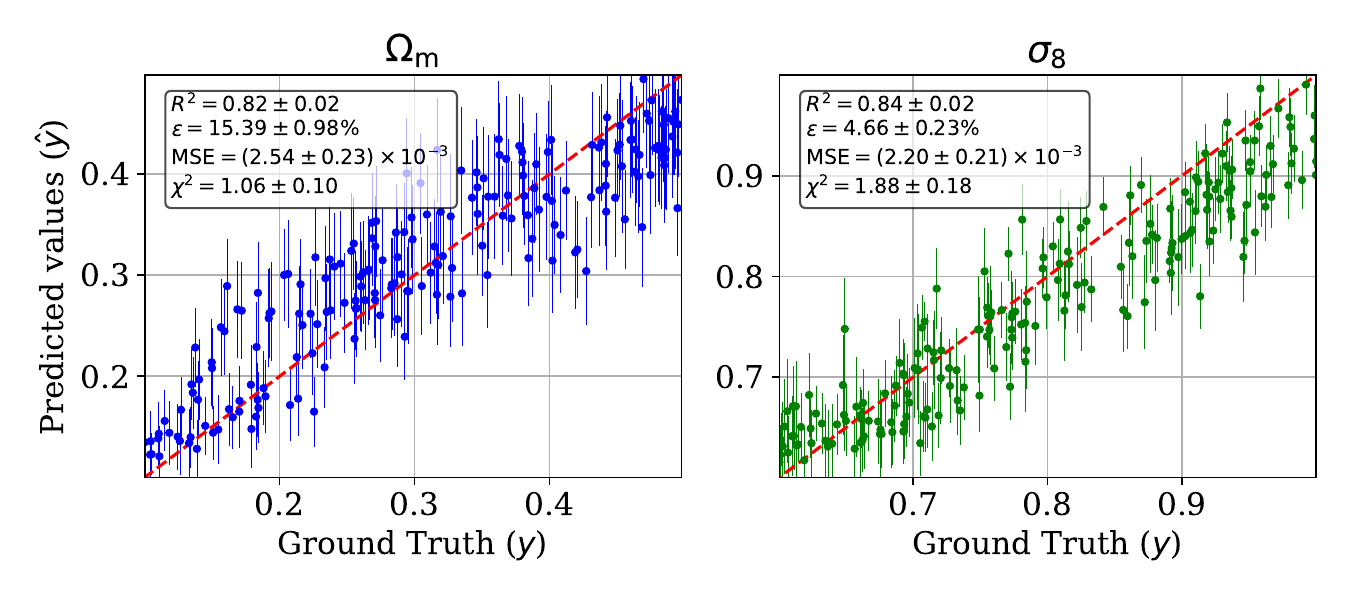}
    \caption{\label{fig:best_result} 
      The evaluation results for our most effective model on the Quijote suite, ClusterTNN, derived from its \IsolatedRun{} are presented. Accuracy metrics with bootstrapped standard deviations on the test set for \Om{} (\textit{left, blue}) and \se{} (\textit{right, green}) are depicted. ClusterTNNs exhibit statistically significant improvements over GNNs, with gains of 22\% for \Om{} and 34\% for \se{}. Our topological neural network outperforms the point cloud-based neural networks employed in \cite{Chatterjee2024} in both \Om{} and \se{} cases. Additionally, it demonstrates superior performance on \se{} compared to any other model utilizing GNNs and two-point correlation functions (2PCFs) examined in \cite{MIT_Benchmark}. For a comprehensive comparative analysis, refer to Table \ref{table:summary_quijote} and Section \ref{sec:quijote_results}.
    }
    \vspace{2mm}
    \end{figure*}
    
\subsection{\label{sec:training}Training Procedure}

A TNN is characterized by a series of parameters, including the number and type of layers. The performance of the model also depends on additional parameters, such as the learning rate, weight decay, and linking radius $r_{\rm link}$, among others. To identify the model that performs the best, we perform hyperparameter optimization using Optuna \cite{optuna}. For each \IsolatedRun{}, we conducted 100 trials for both CAMELS and Quijote. For \IntegratedRun{}, we performed 300 tests for CAMELS and Quijote.

For Quijote catalogs, we set the maximum number of tetrahedra at $N_{\rm cut} \in \{3\text{\small,}000, 4\text{\small,}000, 5\text{\small,}000\}$. This limitation is necessary because our machines cannot process all data effectively with too many tetrahedra due to insufficient memory. In these catalogs, we also set $r_{\rm link}=0.015$. In contrast, for CAMELS, the linking radius is chosen among $r_{\rm link} \in \{0.010, 0.015, 0.020\}$. This discrepancy in the methodology for creating the combinatorial complexes stems from the distinct characteristics of the two datasets. The Quijote catalogs comprise 5\text{\small,}000 halos, which necessitates cuts in the number of cells, whereas CAMELS typically contains fewer than 1,000 galaxies per catalog. Unlike \cite{CosmoGraphNet}, we vary $r_{\rm link}$ in discrete increments, as reconfiguring cells and neighborhood matrices on the fly for TNNs is significantly more computationally costly than for a GNN. 

We reserve 80\% of the samples for training, 10\% for validation, and 10\% for testing. For more information on the complete list of hyperparameters, refer to Appendix \ref{sec:hyperparam}.

\begin{deluxetable*}{l|cccc|cccc}[t]
\tablewidth{0pt} 
\label{table:summary_quijote}
\tablecaption{Comparison of diverse neural network architectures' performance on the Quijote suite.}
\tablehead{
\colhead{} & 
%\colhead{} &
\multicolumn{4}{c}{\Om{}} &
\multicolumn{4}{c}{\se{}} \\
\cline{2-5} \cline{6-9}
%\colhead{Category/Reference} & 
\colhead{Architecture} & 
\colhead{$R^{2}$} & 
\colhead{$\epsilon$(\%)} & 
\colhead{$\chi^{2}$} & 
\colhead{MSE} & 
\colhead{$R^{2}$} & 
\colhead{$\epsilon$(\%)} & 
\colhead{$\chi^{2}$} & 
\colhead{MSE}}

\startdata
 GNN 
 & $\valpm{0.76}{0.03}$ 
 & $\valpm{15.83}{0.77}$ 
 & $\valpm{2.42}{0.58}$ 
 & $\valpm{3.26}{0.31}$ 
 & $\valpm{0.77}{0.03}$ 
 & $\valpm{6.20}{0.35}$ 
 & $\valpm{3.43}{0.31}$ 
 & $\valpm{3.31}{0.34}$
 \\
 TetraTNN & $\valpm{0.78}{0.03}$ 
 & $\mathbf{\valpm{15.33}{0.84}}$ 
 & $\valpm{1.39}{0.13}$ 
 & $\valpm{3.07}{0.31}$ 
 & $\valpm{0.84}{0.02}$ 
 & $\valpm{4.97}{0.29}$ 
 & $\valpm{2.16}{0.24}$ 
 & $\valpm{2.31}{0.26}$
 \\
ClusterTNN 
& $\mathbf{\valpm{0.82}{0.02}}$ 
& $\valpm{15.39}{0.98}$ 
& $\mathbf{\valpm{1.06}{0.10}}$ 
& $\valpm{2.54}{0.23}$ 
& $\valpm{0.84}{0.02}$ 
& $\valpm{4.66}{0.23}$ 
& $\valpm{1.88}{0.18}$ 
& $\valpm{2.20}{0.21}$ 
\\
FullTNN 
& $\valpm{0.80}{0.02}$ 
& $\valpm{16.01}{0.96}$ 
& $\valpm{1.57}{0.16}$ 
& $\valpm{2.79}{0.23}$ 
& $\valpm{0.89}{0.02}$ 
& $\valpm{3.87}{0.25}$ 
& $\mathbf{\valpm{1.08}{0.16}}$ 
& $\valpm{1.55}{0.25}$
\\
\cline{2-9} 
FullTNN (\IntegratedRun{}) 
& $\valpm{0.79}{0.02}$ 
& $\valpm{15.34}{0.80}$ 
& $\valpm{1.39}{0.12}$ 
& $\valpm{2.89}{0.26}$ 
& $\mathbf{\valpm{0.91}{0.01}}$ 
& $\mathbf{\valpm{3.69}{0.21}}$ 
& $\valpm{1.67}{0.25}$ 
& $\mathbf{\valpm{1.33}{0.17}}$ 
\\
\hline
PointMLP-elite variant 
& $\phantomvalpm{0.80}{0.00}$ 
& $\phantomvalpm{15.6}{0.000}$
& $\phantomvalpm{0.81}{0.00}$
& $\phantomvalpm{2.8}{0.000}$ 
& $\phantomvalpm{0.01}{0.000}$
& $\phantomvalpm{13.4}{0.00}$
& $\phantomvalpm{1.25}{0.00}$
& $\phantomvalpm{14}{0.0000}$ 
\\
\hline
 2PCF & & & & $\valpm{2.03}{0.02}$ & & & & $\valpm{4.66}{0.06}$ \\
 GNN & & & & $\valpm{2.77}{0.41}$ & & & & $\valpm{4.84}{2.90}$ \\
 EGNN & & & & $\valpm{13.33}{0.00}$ & & & & $\valpm{13.37}{0.00}$ \\
 NequIP ($\ell_{\rm max} = 1$) & & & & $\valpm{2.88}{0.15}$ & & & & $\valpm{5.05}{1.08}$ \\
 NequIP ($\ell_{\rm max} = 2$) & & & & $\valpm{3.07}{0.18}$ & & & & $\valpm{4.80}{0.49}$ \\
 SEGNN ($\ell_{\rm max} = 1$) & & & & $\valpm{2.31}{0.03}$ & & & & $\valpm{2.34}{0.08}$ \\
 SEGNN ($\ell_{\rm max} = 2$) & & & & $\valpm{2.37}{0.06}$ & & & & $\valpm{2.36}{0.22}$ \\
 PointNet++ & & & & $\valpm{2.87}{0.07}$ & & & & $\valpm{9.00}{3.94}$ \\
 SEGNN ($\ell_{\rm max} = 2$) $+$ 2PCF & & & & $\mathbf{\valpm{1.66}{0.01}}$ & & & & $\valpm{2.38}{0.07}$ \\
 SEGNN ($\ell_{\rm max} = 2$) $+$ 2PCF$_{\rm small}$ & & & & $\valpm{2.27}{0.01}$ & & & & $\valpm{2.40}{0.04}$ \\
 SEGNN ($\ell_{\rm max} = 2$) $+$ 2PCF$_{\rm large}$ & & & & $\valpm{1.73}{0.04}$ & & & & $\valpm{2.26}{0.09}$ 
\enddata
\tablecomments{
The table presents a comparative analysis of various neural network architectures on the Quijote suite, focusing on their predictive performance for two cosmological parameters: \Om{} and \se{}. The metrics include the coefficient of determination ($R^2$), the relative error ($\epsilon$), the chi-squared error ($\chi^2$), and the mean squared error (MSE, reported in units of $10^{-3}$), along with standard deviations of the metrics, derived from bootstrapped estimates on the test set. The \IntegratedRun{} finds the FullTNN as the best model on the validation set. We follow our results with a comparison to existing benchmarks, including the PointMLP-elite variant (\cite{Chatterjee2024}) and several models from \cite{MIT_Benchmark}. The bold values highlight the best performance for each metric.}
\end{deluxetable*}

\subsection{\label{sec:loss}Loss Function}
For a given combinatorial complex $(S, \chi, \text{rk})$, we employ TNNs to estimate the marginal posterior mean $\mu_{i}$ (first moment) and the standard deviation $\sigma_{i}$ (second moment) of the cosmological parameters. Our neural networks are trained to predict \Om{} and \se{} for Quijote, as well as \Om{} for CAMELS,\footnote{We have tried to infer the value of $\sigma_8$ with TNNs for CAMELS but we didn't find any model that yield accurate results.} aligning with established benchmarks \citep{MIT_Benchmark, Chatterjee2024, CosmoGraphNet}. The TNNs accept a combinatorial complex as input and produce the vector $\mathbf{y}=[\mu_i, \sigma_i]$, with each component defined by,
\begin{eqnarray}
    \mu_i(S, \chi, \text{rk}) = \int_{\theta_i} p(\theta_i |S, \chi, \text{rk})\theta_i d\theta_i
\end{eqnarray}
\begin{eqnarray}
    \sigma_i^2(S, \chi, \text{rk}) = \int_{\theta_i} p(\theta_i |S, \chi, \text{rk})(\theta_i - \mu_i)^2 d\theta_i
\end{eqnarray}
where the marginal posterior over the cosmological parameter of interest $\theta_i$ is formulated as,
\begin{eqnarray}
    p(\theta_i |S, \chi, \text{rk}) = \int_{\theta_i} p(\theta_1, \theta_2, \dots, \theta_n |S, \chi, \text{rk})\prod_{j\neq i} d\theta_j.
\end{eqnarray} 

The above quantities are computed by minimizing the following loss function \citep{Jeffrey2020}:
\begin{eqnarray} \label{eqn:loss}
    \mathcal{L}_{i}{=}\frac{1}{B} \left ( \sum_{j \in \mathcal{B}} (\theta_{i,j}-\mu_{i,j})^2{+}
    \sum_{j \in \mathcal{B}} ((\theta_{i,j}{-}\mu_{i,j})^2 {-}\sigma_{i,j})^2 \right)
\end{eqnarray}
where $B$ represents the size of the batch $\mathcal{B}$. The total loss is obtained by summing over all the cosmological parameters of interest: $\mathcal{L}=\sum_{i} \mathcal{L}_i$. We empirically found that the original form of simply adding the two terms in Equation \ref{eqn:loss} performs better in predicting the standard deviations than alternative methods of adding the logarithmically scaled terms \citep{CosmoGraphNet,Shao2023,deSanti2023,Chatterjee2024}.

Here, the standard deviation $\sigma_i$ represents the aleatoric error, which is the statistical error arising from cosmic variance. It does not account for the epistemic error, or the error associated with individually trained neural networks. In this study, we focus solely on the aleatoric error, as our aim is to compare the differences in the constraining power that arise from incorporating topologies into neural networks, alongside other benchmarks. Moreover, previous works have identified the magnitude of the epistemic error as significantly smaller than the aleatoric error when working with GNNs \citep{deSanti2023}. Therefore, we also anticipate it to be negligible in the case of TNNs.

\subsection{\label{sec:metrics}Validation Metrics}
Throughout our analysis, we employ the standard set of accuracy metrics as defined in \cite{CosmoGraphNet, Natali_2023} to measure each cosmological parameter within the test set:

\begin{enumerate}
    \item \textbf{Coefficient of determination} ($R^2$)
        \begin{eqnarray}
            R^2 = 1- \frac{\sum_{i} (\theta_i-\mu_i)^2}{\sum_{i} (\theta_i-\bar{\theta})^2}
        \end{eqnarray}
    \item \textbf{Mean relative error} ($\epsilon$)
        \begin{eqnarray}
            \epsilon = \frac{1}{N} \sum_{i} \frac{|\theta_i-\mu_i |}{\theta_i}
        \end{eqnarray}
    \item \textbf{Chi squared} ($\chi^2$)
        \begin{eqnarray}
            \chi^2 = \frac{1}{N} \sum_{i} \frac{(\theta_i-\mu_i)^2}{\sigma_i^2}
        \end{eqnarray}
    \item \textbf{Mean squared error} (MSE)
        \begin{eqnarray}
            \text{MSE} = \frac{1}{N} \sum_{i} (\theta_i-\mu_i)^2
        \end{eqnarray}
\end{enumerate}
A precise and accurate model will exhibit low values of $\epsilon$ and MSE, while demonstrating high values of $R^2$ and $\chi^2\simeq1$.

\begin{deluxetable*}{l|cccc}[t]
\tablewidth{0pt} 
\label{table:summary_camels}
\tablecaption{Comparison of diverse neural network architectures' performance on the CAMELS suite.}
\tablehead{
\colhead{} &
\multicolumn{4}{c}{\Om{}} \\
\cline{2-5}
\colhead{Architecture} & 
\colhead{$R^{2}$} & 
\colhead{$\epsilon$(\%)} & 
\colhead{$\chi^{2}$} & 
\colhead{MSE}}

\startdata
GNN 
& $\valpm{0.76}{0.04}$ 
& $\valpm{14.02}{1.12}$ 
& $\valpm{1.51}{0.19}$  
& $\valpm{3.05}{0.42}$ 
\\
TetraTNN 
& $\valpm{0.81}{0.04}$ 
& $\mathbf{\valpm{13.09}{1.20}}$ 
& $\valpm{1.54}{0.25}$  
& $\mathbf{\valpm{2.41}{0.46}}$ 
\\
ClusterTNN 
& $\valpm{0.67}{0.07}$ 
& $\valpm{16.79}{1.56}$ 
& $\valpm{6.80}{1.29}$  
& $\valpm{4.14}{0.73}$ 
\\
FullTNN 
& $\valpm{0.75}{0.05}$ 
& $\valpm{14.98}{1.34}$ 
& $\valpm{7.43}{1.28}$  
& $\valpm{3.21}{0.51}$ 
\\
\cline{2-5}
GNN (\IntegratedRun{}) & $\valpm{0.77}{0.04}$ & $\valpm{14.34}{1.22}$ & $\valpm{1.65}{0.23}$ & $\valpm{2.85}{0.43}$\\
\hline
GNN (\cite{CosmoGraphNet}) 
& $\mathbf{\phantomvalpm{0.83}{0.000}}$
& $\phantomvalpm{13.1}{0.000}$ 
& $\mathbf{\phantomvalpm{1.24}{0.00}}$ 
& 
\enddata
\tablecomments{
The table presents a comparative analysis of various neural network architectures on the CAMELS suite, focusing on their predictive performance for \Om{}. The metrics include the coefficient of determination ($R^2$), the relative error ($\epsilon$), the chi-squared error ($\chi^2$), and the mean squared error (MSE, reported in units of $10^{-3}$). The results are categorized by the type of training run (\IsolatedRun{} or \IntegratedRun{}) and compared with existing methods. The bold values highlight the best performance for each metric.}
\end{deluxetable*}

\section{Results}
\label{sec:results}
In this section we describe the results we obtain by training TNNs on data from both the Quijote and CAMELS simulations. We report the results obtained from the two different setups: \IsolatedRun{} and \IntegratedRun{}. A detailed summary of the results we obtain is provided in Table \ref{table:summary_quijote} for Quijote and Table \ref{table:summary_camels} for CAMELS, together with previous results by other groups. The listed test results are from the best models, selected based on their performance on the validation set across multiple trials in the \Optuna{} study. Furthermore, we report the mean and standard deviation for all metrics by bootstrapping from the test set, enabling a clear comparison of model performance. We also note that the train, validation, and test splits are identical across all setups.\footnote{A much more thorough error analysis would involve bootstrapping over multiple random data splits, which is computationally prohibitive in our case. However, the inclusion of a validation set helps mitigate overfitting to the training data, and we ensure consistent comparisons by using identical splits across all models.}

\subsection{Quijote \label{sec:quijote_results}}
The overall training results are presented in Table \ref{table:summary_quijote}, which also includes findings from previous research. We highlight in bold the best values obtained. From the \IntegratedRun{}, we identify FullTNN as the best-performing model, particularly in \se{}---up to 60\% improvement in terms of MSE. Within the \IsolatedRun{}, ClusterTNNs achieve the best performance in \Om{} (with the exception of relative error), while FullTNNs perform best in \se{}. Figure \ref{fig:best_result} depicts the detailed test set results for our best model. 

Specifically, comparing \IsolatedRun{} of ClusterTNNs with GNNs, we observe an overall improvements in MSE of 22\% for \Om{} and 34\% for \se{}. FullTNNs also demonstrate improvements of 14\% in \Om{} and a substantial gain of 53\% in \se{}. TetraTNNs, the most similar architecture to GNNs among the TNNs, exhibit moderate improvements in both \Om{} and \se{}, with gains of 6\% and 30\%, respectively. Overall, our TNNs augmented with higher-order message-passing demonstrate clear superiority over the standard GNN for both cosmological parameters, even when accounting for bootstrapped uncertainties. Furthermore, the reduced uncertainties observed in the ClusterTNN and FullTNN models indicate that their results are more reliable and robust. Although FullTNNs are intended to fully represent TNNs, their improvement in performance for \Om{}, when jointly predicted with \se{}, is comparable but less significant. This may arise due to the fact that FullTNNs may require longer trials to achieve a similar level of optimization due to their inherent complexity.

When we compare our TNNs against the recent study of \cite{Chatterjee2024} that uses a variant of PointMLP-elite \citep{PointMLP-elite}, we observe that the absence of higher-order message-passing is critical, especially for constraining \se{}. Given that we achieve better values for every metric even though we work with fewer halos (5\text{\small,}000 vs. 8\text{\small,}192), we demonstrate the importance of incorporating higher-order message-passing structures. This conclusion is also supported by the PointNet++ architecture utilized in \cite{MIT_Benchmark}. A study by \cite{DiffusionPointCloud}, which employs point cloud and diffusion-based generative models, claims to achieve promising results of $\epsilon\approx5\%$ and $\epsilon\approx3\%$ for \Om{} and \se{}, respectively. Despite these findings, their study indicates that the likelihoods of the model are not properly calibrated, showing an overconfidence in \Om{}, as demonstrated in Figure 7 of their paper. Furthermore, they report that their GNNs show inferior predictions on the likelihoods relative to their point cloud-based neural networks, contrary to the usual trend in which GNNs often surpass point cloud-based neural networks, as illustrated in Table \ref{table:summary_quijote}.

Compared to various GNN approaches in \cite{MIT_Benchmark}, our method demonstrates a similar level of accuracy for \Om{} and a marked improvement for \se{}. Compared to the best-performing GNN model that omits the direct use of the two-point correlation function (2PCF), SEGNN ($l_{\rm max }=1$), our FullTNN model from \IntegratedRun{} provides a 43\% improvement in MSE for \se{}, while only experiencing a 25\% increase in MSE for \Om{}. Even for approaches that explicitly integrate the 2PCF information, our model yields better MSE values, showing an improvement of 41\% in \se{}. In summary, our models considerably improve the constraints in \se{} while maintaining comparable performance in \Om{}. Furthermore, it is important to note that, unlike \cite{MIT_Benchmark}, which focuses on minimizing a single MSE metric through the MSE loss function, our approach also predicts the standard deviation for each parameter.

We caution the reader that all of these comparisons should be viewed with a grain of salt. In the case of \cite{Chatterjee2024}, while the data is comparable, the number of halos differs. \cite{MIT_Benchmark} employs halo catalogs identified by the {\rm R{\scriptsize OCKSTAR}} halo finder \citep{Rockstar}, in contrast to our use of the FoF method.\footnote{We perform additional tests using halo catalogs identified by the {\rm R{\scriptsize OCKSTAR}} halo finder and observe a degradation in performance compared to those from the FoF halo finder. We plan to investigate the robustness of different TNN architectures across various halo finders in future work.} Additionally, they train their models on a larger number of simulations that use Sobol sequences to uniformly sample the parameter space. However, in our case, while all simulations are generated using latin-hypercube sampling, once we split the dataset into training, validation, and testing sets, the training set may not maintain uniform coverage of the parameter space.

\subsection{CAMELS\label{sec:camels_results}}
We evaluate our quartet of models on the CAMELS dataset by conducting an analysis similar to that used for the Quijote suite. As illustrated in Table \ref{table:summary_camels}, contrary to the findings of the Quijote suite, GNNs are preferred over other TNNs for the \IntegratedRun{}, and TetraTNNs show the best results among the \IsolatedRun{}s. The results reveal an inverse pattern, indicating that the added complexity in the architecture diminishes the performance for predicting \Om{}. Additionally, TetraTNN, which was previously outperformed by other TNNs in the Quijote suite scenario, actually excels in the CAMELS suite. Specifically, the \IsolatedRun{} for TetraTNNs achieves a 21\% reduction in MSE for \Om{} compared to GNNs.
 Compared to previous research using a GNN \citep{CosmoGraphNet}, our top outcome of \IntegratedRun{} demonstrates a slightly degraded performance across all metrics. However, we emphasize that we were able to reproduce the comparable level of performance using a similar architecture, albeit with a completely different framework.\footnote{We note that, due to the heavy computational cost of generating combinatorial complexes, we coarsely sample the hyperparameter space for $r_{\rm link}$, unlike in \cite{CosmoGraphNet} (see Appendix \ref{sec:hyperparam} for details). A finer sampling of $r_{\rm link}$—which controls the sparsity of the graph—could potentially lead to improved results.}

Moreover, we observe that although higher-order cells and message-passing are incorporated through our TNNs, we were unable to effectively constrain \se{} for the CAMELS suite. This outcome contrasts with the Quijote suite, where TNNs significantly improved \se{} performance compared to GNNs. We hypothesize that the limited number of galaxies, the complexities introduced by hydrodynamics, supernova and AGN feedback, and the smaller box size of the CAMELS suite pose challenges for accurate \se{} prediction by the machines. We note that the gradual addition of higher-order structures, from edges to clusters, appears to diminish performance in the CAMELS suite compared to the Quijote suite. In particular, incorporating clusters diminishes the performance more than incorporating tetrahedra unlike the Quijote suite. This suggests that the majority of the cosmological information likely originates from smaller spatial scales, and GNNs may already have a saturated information gain from the CAMELS simulations. Therefore, while theoretically TNNs extend GNNs, in practice it proves challenging to optimize using information from higher-order interactions and larger scales.

We note that there is a clear difference between the Quijote and CAMELS catalogs in terms of the number of objects they contain. Although the Quijote catalogs consistently include 5\text{\small,}000 objects, the number in the CAMELS catalogs varies from one catalog to another. We believe that some information may be associated with the number density; therefore, this variation could potentially underestimate the power of TNNs. Unfortunately, maintaining a consistent number density in the CAMELS catalogs is not feasible, as some catalogs contain only a few galaxies. However, with the second-generation CAMELS simulations, this challenge may be addressed, making it possible to repeat the analysis using galaxy catalogs with a fixed number density.

\section{Conclusions}
\label{sec:conclusions}
In this work, we introduce topological neural networks (TNNs) as powerful architectures for field-level inference on cosmological data. Known for their enhanced expressiveness, performance, and efficiency in long-range message-passing, we applied these novel architectures to cosmological and astrophysical datasets for the first time, to the best of our knowledge. Using the $N$-body simulation suite, Quijote, and the (magneto-)hydrodynamic suite CAMELS, we investigate how TNNs perform compared to existing benchmarks, employing point cloud-based neural networks and GNNs. 

We first build combinatorial complexes, which are powerful mathematical structures to abstract both hierarchies and flexibility among higher-order cells. Through convolutional push-forward and merge node operations, we enable message-passing between different neighboring cells defined via incidence and (co)-adjacencies  (Section \ref{sec:message-passing}). Specifically, we define not only individual vertices and edges, but also higher order cells: tetrahedra using Delaunay triangulation, clusters using HDBSCAN, and hyperedges identified through the minimum spanning tree edges connecting neighboring clusters (Section \ref{sec:topology_cosmic_data}). The growing hierarchy, from vertices to hyperedges, serves as a proxy for abstracting cosmological information residing at different scales. Our TNNs also respect the $E(3)$-invariance concerning translations, reflections, and rotations. This is accomplished by defining not only scalars such as numbers, angles, lengths, areas, or volumes assigned to individual cells but also scalars involving two heterogeneous cells participating in message-passing (Section \ref{sec:symmetry}). 

In order to observe the effect of adding topologies arising from different ranks of cells, we compare four architectures with a gradual increase in complexity: GNN (vertices and edges), TetraTNN (vertices, edges, and tetrahedra), ClusterTNN (vertices, edges, and clusters) and FullTNN (vertices, edges, tetrahedra, clusters, and hyperedges). To test the full capabilities of our models, we enhance their flexibility by incorporating various tunable hyperparameters. We conduct two flavors of runs, \IsolatedRun{} that evaluate the performance of each of the architectures, and \IntegratedRun{} that allows the optimizer to select from the four architectures (see Section \ref{sec:training} and Appendix \ref{sec:hyperparam}). Below are the key takeaways from our work.

\begin{itemize}
    \item Our TNNs demonstrate a striking improvement in constraining \se{} from the Quijote suite, particularly excelling where point cloud-based architectures fail completely. Compared to our GNNs, ClusterTNNs show improvements with up to 22\% in \Om{} and 34\% in \se{}, while the best-performing FullTNN achieves an improvement of up to 60\% in \se{}.
    \item When applied to the CAMELS suite, GNNs outperform other TNNs, except for TetraTNNs, in constraining \Om{}. Our GNNs perform comparable to the existing benchmark despite differences in implementation and subtle architectural variations.
    
    \item For Quijote, the gradual incorporation of higher-order structures proves effective, whereas the CAMELS suite shows an inverse relationship. We speculate that this difference arises from the lack of subhalos and small box volumes in CAMELS, which limit the expressive power gained from adding higher-order structures.
    
    \item Our finding that TNNs do not aid in constraining \se{} from CAMELS---unlike Quijote---suggests that GNNs already saturate the available information from simulations, while TNNs are excessive and more challenging to train. Additionally, unlike in the Quijote suite, tetrahedral cells are more advantageous than clusters, implying that most of the relevant information resides at smaller scales.
\end{itemize}

We have explored that, depending on the dataset, TNNs are effective in extracting cosmological information related to topologies. We highlight that our development of combinatorial complexes involves ad hoc engineering and offers significant potential for further investigation. For example, it is possible to accurately model the large-scale structure by identifying galaxy superclusters or filaments instead of clusters or hyperedges used in this work. More sophisticated measures of $E(3)$-invariant features can also be introduced to improve performance. Furthermore, our architectures and the definition of neighborhoods could be relaxed for further optimization. Although our TNNs were designed for complex-level predictions (i.e. cosmological parameters), our analysis can be extended to include not only galaxy or halo-level predictions but also predictions concerning various types of large-scale structures. Another area of investigation is the robustness of TNNs relative to GNNs when utilizing simulations run with alternative subgrid physics models, especially for the CAMELS suite. Alongside these avenues for improvement, we also plan to extend our evaluation of TNNs to cosmological simulations beyond $\Lambda$CDM in future work.

\section*{Acknowledgments} % For Arxiv Submission
%\begin{acknowledgments}
We thank Adrian Bayer, Teresa Huang, Ji-hoon Kim, Natali de Santi, Lawrence Saul, and Soledad Villar for their valuable discussions. Jun-Young Lee’s work was supported by the Global-LAMP Program of the National Research Foundation of Korea (NRF) grant, funded by the Ministry of Education (No. RS-2023-00301976).  His work was also supported by the National Institute of Supercomputing and Network/Korea Institute of Science and Technology Information, with supercomputing resources including technical support, under grants KSC-2020-CRE-0219, KSC-2021-CRE-0442, and KSC-2022-CRE-0355. We also acknowledge the use of the local GPU clusters {\rm H{\scriptsize APPINESS}} and {\rm H{\scriptsize ERCULES}} of Seoul National University during the initial phase of this work. The work of FVN is supported by the Simons Foundation. The Flatiron Institute is supported by the Simons Foundation. 

%\end{acknowledgments}

\bibliography{sample631}{}
\bibliographystyle{aasjournal}

\appendix

\begin{deluxetable}{l l l}
\tablecaption{Full List of Hyperparameters}
\label{tab:hyperparams}
\tablehead{
\colhead{Hyperparameter} & \colhead{Description} & \colhead{Values} 
}
\startdata
\multicolumn{3}{c}{\textbf{Data-Related}} \\
\hline
\texttt{data\_mode} (Quijote) & Maximum number of tetrahedra ($N_{\text{tetra}}$) & $\{3\text{\small,}000, 4\text{\small,}000, 5\text{\small,}000\}$ \\
\texttt{data\_mode} (CAMELS)  & Linking radius criterion for connecting two vertices with an edge ($r_{\rm link}$) & $\{0.010, 0.015, 0.020\}$ \\
\texttt{cci\_mode} & $E(3)$-invariant distance metrics between two neighboring cells & \{\texttt{Euclidean, Hausdorff}\}\\
\texttt{drop\_prob} & Drop probability for disconnecting neighbors & [0, 0.2] \\
\hline
\multicolumn{3}{c}{\textbf{Model Architecture-Related}} \\
\hline
\texttt{layer\_type} & Type of the model layers & \{GNN, TetraTNN, ClusterTNN, FullTNN\} \\
\texttt{num\_layers} & Number of TNN or GNN layers & \{1, 2, 3, 4, 5, 6\} \\
\texttt{hidden\_dim} & Dimension of hidden layers & \{32, 64, 128, 256\} \\
\texttt{aggr\_func} & Aggregation function for the model & \{\texttt{sum, max, min, all}\} \\
\hline
\multicolumn{3}{c}{\textbf{Learning-Related}} \\
\hline
\texttt{learning\_rate} & Learning rate used in gradient updates & $[10^{-5}, 10^{-2}]$ \\
\texttt{weight\_decay} & Weight decay parameter (L2 regularization) & $[10^{-5}, 10^{-3}]$ \\
\texttt{batch\_size} & Batch size used for gradient accumulation & \{1, 2, 4, 8\} \\
\texttt{T\_max} & Maximum number of epochs for the cosine annealing scheduler & [10, 100] \\
\texttt{update\_func} & Activation function used for updates & \{\texttt{ReLu, Tanh}\} \\
\enddata
\tablecomments{
The table presents a full list of hyperparameters together with their descriptions and values. The hyperparameters are categorized into three main groups: data-related, model architecture-related, and learning-related. We employ the \Optuna{} framework to conduct a thorough study within the given scope.}
\end{deluxetable}

\section{\label{sec:hyperparam}Full List of Hyperparameters and Training Details}
In this section, we provide the complete training details of our topological neural networks. As organized in Table \ref{tab:hyperparams}, we categorize our hyperparameters into three groups: data-related, model architecture-related, and learning-related. Given the range of values, the optimizer selects each hyperparameter, \Optuna{} with the Tree-structured Parzen Estimator algorithm \citep{optuna}. As explained in Section \ref{sec:training}, the \texttt{data\_mode} for Quijote focuses on the number of tetrahedra selected. The tetrahedra are sorted in ascending order of volume. For CAMELS, following the practice of \cite{CosmoGraphNet}, we vary the linking radius, which sets a criterion for connecting two vertices with an edge. The \texttt{cci\_mode} indicates the distance metric for the cell-cell $E(3)$-invariance, and we use Euclidean or Hausdorff metrics as explained in Section \ref{sec:symmetry}. We also introduce a feature of randomly dropping neighbors before training to observe the effect of varying connectivity by controlling the probability \texttt{drop\_prob}. 

With respect to the hyperparameters associated with model architectures, we explore variations in layer types, layer quantities, and aggregation methods. It has been suggested that the inclusion of various neighborhood aggregation functions can improve the effectiveness of GNN \citep{PNA}. Within our TNNs, three distinct aggregation operations are employed: convolutional push-forward, convolutional merge node, and global pooling. For each of these operations, we utilize multiple aggregation functions. For global pooling, we apply a predetermined set of average, standard deviation, maximum, and minimum aggregation functions. For both convolutional push-forward and merge node operations, we select from \{\texttt{sum, max, min, all}\}. The \texttt{all} aggregation mode includes \{\texttt{sum, max, min}\} operations and appends the \texttt{std} aggregation exclusively to the intra-neighborhood aggregation for the merge node operations. This distinction is due to the possibility that certain cells might not have any neighboring cells. When using multiple aggregation functions, they undergo an additional multilayer perceptron layer and an activation function to maintain dimensionality.

Finally, we set the learning-related parameters \texttt{learning\_rate}, \texttt{weight\_decay}, \texttt{batch\_size}, \texttt{T\_max}, and \texttt{update\_func}. Our pipeline currently does not support explicit batching, but utilizes gradient accumulation to mimic the effect of batching. Furthermore, our learning rate is set using the CosineAnnealingLR scheduler \citep{CosineAnnealingLR}. We refer the reader to our \githubrepo{https://github.com/Byeol-Haneul/CosmoTopo}{Github repository} to explore the details of our neural network.

%% For this sample we use BibTeX plus aasjournals.bst to generate the
%% the bibliography. The sample631.bib file was populated from ADS. To
%% get the citations to show in the compiled file do the following:
%%
%% pdflatex sample631.tex
%% bibtext sample631
%% pdflatex sample631.tex
%% pdflatex sample631.tex

%% This command is needed to show the entire author+affiliation list when
%% the collaboration and author truncation commands are used. It has to
%% go at the end of the manuscript.
%\allauthors

%% Include this line if you are using the \added, \replaced, \deleted
%% commands to see a summary list of all changes at the end of the article.
%\listofchanges
\end{document}